\documentstyle[aps,epsfig,prl,graphicx,twocolumn]{revtex}

\begin{document}
\title{Josephson and spontaneous currents at the interface between two $d$-wave
superconductors with transport current in the banks}
\author{Yu.A. Kolesnichenko, A.N. Omelyanchouk, S. N. Shevchenko}
\address{B.I. Verkin Institute for Low Temperature Physics and Engineering, National\\
Academy of Sciences of Ukraine, 47 Lenin Ave., 61103, Kharkov, Ukraine}
\maketitle

\begin{abstract}
A stationary Josephson effect in the ballistic contact of two
d-wave superconductors with different axes orientation and with
tangential transport current in the banks is considered
theoretically. We study the influence of the transport current on
the current-phase dependence for the Josephson and tangential
currents at the interface. It is demonstrated that the spontaneous
surface current at the interface depends on the transport current
in the banks due to the interference of the angle-dependent
current-carrying condensate wave functions of the two
superconductors.
\end{abstract}


\section{Introduction}

It was shown that in the ground state of the contact of two $d$-wave
superconductors with different axes orientation there is the tangential to
the boundary current \cite{Yip}-\cite{Ilichev}. In the particularly
interesting case of $\pi /4$-misorientation the ground state is two-fold
degenerate: there are the tangential currents in opposite directions at $%
\phi =\pm \pi /2$ in the absence of Josephson current. The probabilities to
find the contact in one of the two states are equal and the corresponding
tangential current is referred to as the spontaneous one. It was proposed to
use such two-state quantum systems for the quantum computation \cite{Ioffe}-
\cite{Amin2003}. It is of interest to study the possibility to control this
system by the external transport current, which is the motivation of the
present work.

In the described problem of the Josephson contact of two $d$-wave
superconductors with transport current in the banks the resulting tangential
current is not a sum of the spontaneous current and the transport one. In
the paper \cite{K0Sh(2003)} we have studied simpler case of the contact of
two $s$-wave superconductors with the transport current flowing in the
banks. It was shown that the presence of the magnetic field \cite{Heida}-
\cite{AOZ2}, of the transport superconducting current \cite{K0Sh(2003)}, or
of the current in normal layer \cite{Morpurgo}-\cite{Wilhelm} in a
mesoscopic Josephson junction can significantly influence current-phase
characteristics, current distribution etc.

In the present problem the Josephson current is defined by the interference
of the angle-dependent condensate wave functions of the two superconductors.
There are two factors of anisotropy which define the angle dependence of the
order parameter: the pairing anisotropy and the transport current. Thus, it
is natural to expect that the resulting interference current (which has both
normal and tangential components) is parametrized by the external phase
difference $\phi $ and by the value of the transport current (or by the
superfluid velocity $v_{s}$). The presence of these two controlling
parameters can be useful in the applications of Josephson junctions of high-T%
$_{c}$ superconductors.

In Sec.2 we derive basic equations to describe the ballistic planar
Josephson junction of two differently orientated $d$-wave superconductors
with homogeneous current in the banks. These equations are solved
analytically in Sec.3. Then we study in Sec.4 the influence of the transport
current on the Josephson current and vice versa at the interface. In
Appendix the order parameter and the current density in the homogeneous
situation are considered.

\section{Model and basic equations}

\begin{figure}[t]
\centering
\includegraphics[width=7 cm]{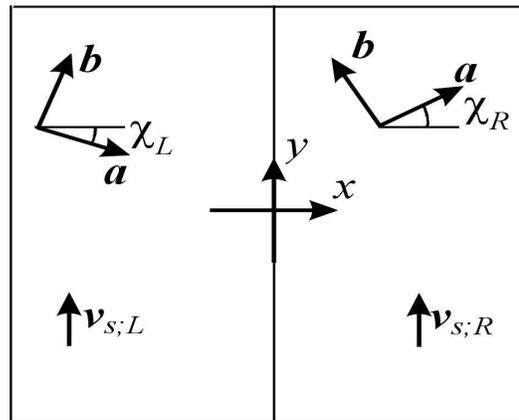}
\caption{Geometry of the contact of two superconductors with
different axes
orientation and different transport currents (superfluid velocities ${\bf v}%
_{s;L,R}$) in the banks.} \label{scheme}
\end{figure}

We consider a model of the Josephson junction as an ideal plane
between two singlet (particularly, $d$-wave) superconductors with
different axes orientation (see Fig.\ref{scheme}).

The pair breaking and the scattering at the junction as well as
the electron scattering in the bulk of metals are ignored. We did
not take into account the possibility of the generation of a
subdominant order parameter, which results in decreasing of the
amplitude of the current \cite{AOZ}. The $c$-axis of both
superconductors is parallel to
the interface. The direction of $c$-axis is chosen as $z$-axis. The $a$ and $%
b$ axes are situated in $xy$-plane. In the banks of the contact homogeneous
current flows with a superconducting velocity ${\bf v}_{s}$. We consider
superfluid velocity ${\bf v}_{s}$ in left (L) and right (R) superconducting
half-spaces being parallel to each other ${\bf v}_{sL}\Vert {\bf v}_{sR}$
and to the boundary; we choose $y$-axis along ${\bf v}_{s}$ and $x$-axis
perpendicular to the boundary; $x=0$ is the boundary plane.

We describe the coherent current state in the superconducting ballistic
structure in the quasiclassical approximation by the Eilenberger equation
\cite{Eilen}, \cite{KO}
\begin{equation}
{\bf v}_{F}\frac{\partial }{\partial {\bf r}}\widehat{G}+\left[ \widetilde{%
\omega }\widehat{\tau }_{3}+\widehat{\Delta },\widehat{G}\right] =0,
\label{Eilenberger Eq}
\end{equation}
where $\widetilde{\omega }=\omega _{n}+i{\bf p}_{F}{\bf v}_{s}$, $\omega
_{n}=\pi T(2n+1)$ are the Matsubara frequencies, $\widehat{G}=\widehat{G}%
_{\omega }({\bf v}_{F},{\bf r})=\left(
\begin{array}{cc}
g & f \\
f^{+} & -g
\end{array}
\right) $ is the energy integrated Green function, $\widehat{\Delta }=\left(
\begin{array}{cc}
0 & \Delta \\
\Delta ^{\ast } & 0
\end{array}
\right) $. The equation (\ref{Eilenberger Eq}) should be supplemented by the
equation for the order parameter (the self-consistency equation):
\begin{equation}
\Delta ({\bf v}_{F},{\bf r})=\pi N_{0}T\sum_{\omega }\left\langle V({\bf v}%
_{F},{\bf v}_{F}^{\prime })f({\bf v}_{F}^{\prime },{\bf r)}\right\rangle _{%
{\bf v}_{F}^{\prime }},  \label{Eq for Delta}
\end{equation}
$N_{0}$ is the density of states at the Fermi level and $\left\langle
...\right\rangle _{{\bf v}_{F}}$ is the averaging over directions of ${\bf v}%
_{F}$; $V({\bf v}_{F},{\bf v}_{F}^{\prime })$ is a pairing attractive
potential. For the bulk d-wave superconductor it is usually assumed $\Delta
(\vartheta )=\Delta _{0}(T,{\bf v}_{s})\cos 2\vartheta $, $V({\bf v}_{F},%
{\bf v}_{F}^{\prime })=V_{d}\cos 2\vartheta \cos 2\vartheta ^{\prime }$,
where angle $\vartheta $ defines a direction of the velocity ${\bf v}_{F}$.
Solutions of Eqs. (\ref{Eilenberger Eq})-(\ref{Eq for Delta}) must satisfy
the conditions for the Green functions and gap function in the banks far
from the interface:
\begin{eqnarray}
g\left( \mp \infty \right) &=&\frac{\omega _{L,R}}{\Omega _{L,R}},
\label{g(inf)} \\
f\left( \mp \infty \right) &=&\frac{\Delta (\mp \infty )}{\Omega _{L,R}},
\label{f(inf)} \\
\Delta (\mp \infty ) &=&\Delta _{L,R}\exp (\pm i\phi /2).  \label{d(inf)}
\end{eqnarray}
Here $\omega _{L,R}=\omega _{n}+i{\bf p}_{F}{\bf v}_{s;L,R},$ $\Omega _{L,R}=%
\sqrt{\omega _{L,R}^{2}+\Delta _{L,R}^{2}}$; $\phi $\ is the phase
difference between the left and right superconductors, which parametrizes
the Josephson current state. The angles $\chi _{L,R}$ define the orientation
of the crystal axes ${\bf a}$ and ${\bf b}$ in left and right half-spaces
(see Fig.\ref{scheme}). The angle between the axes of the right and left
superconductors (the misorientation angle) is $\delta \chi =\chi _{R}-\chi
_{L}$.

Provided we know the Green function $\widehat{G}$, we can calculate the
current density:
\begin{equation}
{\bf j}({\bf r})=-2\pi ieN_{0}T\sum_{\omega }\left\langle {\bf v}_{F}g({\bf v%
}_{F},{\bf r)}\right\rangle _{{\bf v}_{F}}.  \label{j(r)}
\end{equation}

For singlet superconductors it is usually assumed $\Delta (-{\bf v}%
_{F})=\Delta ({\bf v}_{F})$ and therefore we have:
\begin{equation}
f^{+}(\omega ,-{\bf v}_{F})=f^{+}(-\omega ,{\bf v}_{F})=f^{\ast }(\omega ,%
{\bf v}_{F}),
\end{equation}
\begin{equation}
g(\omega ,-{\bf v}_{F})=-g(-\omega ,{\bf v}_{F})=g^{\ast }(\omega ,{\bf v}%
_{F}).  \label{symmetry of g}
\end{equation}
Making use of Eq. (\ref{symmetry of g}), Eq. (\ref{j(r)}) can be rewritten:
\begin{equation}
{\bf j}({\bf r})=-j_{0}\frac{T}{T_{c}}\sum\limits_{\omega >0}\left\langle
\widehat{{\bf v}}_{F}%
\mathop{\rm Im}%
g({\bf r})\right\rangle _{{\bf v}_{F}},~j_{0}=4\pi \left| e\right|
N(0)v_{F}T_{c}.  \label{j}
\end{equation}

\section{Analytical solution of Eilenberger equation}

In this paper we consider the problem non-self-consistently: we assume the
superconducting velocity ${\bf v}_{s}$ being homogeneous and the order
parameter $\Delta $ being constant in the two half-spaces:
\begin{eqnarray}
{\bf v}_{s}({\bf r)} &=&\left\{
{{\bf v}_{s;L},\quad x<0 \atop {\bf v}_{s;R},\quad x>0}%
\right. ,\quad   \label{v_s_and_Delta} \\
\Delta ({\bf r}) &=&\left\{
{\Delta _{L}\exp (i\phi /2),~\quad x<0 \atop \Delta _{R}\exp (-i\phi /2),\quad x>0}%
\right. .  \nonumber
\end{eqnarray}
As it was shown in the papers \cite{AOZ}, the self-consistent consideration
of Josephson junction of $d$-wave superconductors does not qualitatively
differ from the non-self-consistent one. In the paper \cite{AOZ} the authors
compare numerically the self-consistent solution with the
non-self-consistent one. The self-consistency of the solution allows to take
into account the suppression of the order parameter at the interface; the
major effect of this is the reduction of the current \cite{AOZ}.

Eq. (\ref{Eilenberger Eq}) together with Eqs. (\ref{g(inf)}-\ref{d(inf)},
\ref{v_s_and_Delta}) yields for the left and right superconductors:
\begin{equation}
g_{L,R}(x)=\frac{\omega _{L,R}}{\Omega _{L,R}}+C_{L,R}\exp \left( -\frac{%
2\left| x\right| }{\left| v_{x}\right| }\Omega _{L,R}\right) ,  \label{g(x)}
\end{equation}
\begin{eqnarray}
f_{L,R}(x) &=&\frac{\Delta _{L,R}}{\Omega _{L,R}}e^{-%
\mathop{\rm sign}%
(x)i\phi /2}- \\
&&-C_{L,R}\frac{%
\mathop{\rm sign}%
(x)\eta \Omega _{L,R}+\omega _{L,R}}{\Delta _{L,R}}\times   \nonumber \\
&&\times \exp \left( -\frac{2\left| x\right| }{\left| v_{x}\right| }\Omega
_{L,R}\right) e^{-%
\mathop{\rm sign}%
(x)i\phi /2},  \nonumber
\end{eqnarray}
where $\eta =%
\mathop{\rm sign}%
(v_{x})$. Making use of the continuity condition, we obtain the expression
for $g$-function at the interface:
\begin{equation}
g(0)=\frac{\Omega _{L}\omega _{R}+\Omega _{R}\omega _{L}-i\eta \Delta
_{L}\Delta _{R}\sin \phi }{\Omega _{L}\Omega _{R}+\omega _{L}\omega
_{R}+\Delta _{L}\Delta _{R}\cos \phi }.  \label{g(0)}
\end{equation}

Eqs. (\ref{j}) and (\ref{g(0)}) allow us to calculate Josephson current $%
j_{J}=j_{x}(x=0)$ and the tangential current $j_{y}(x=0)$ at the interface.
We underline that these equations are valid to describe the current at the
interface of two singlet superconductors with different axes orientation and
with different transport currents in the banks. The contact of conventional
superconductors was considered in \cite{K0Sh(2003)} and in the present paper
we study the contact of $d$-wave superconductors, for which the order
parameter is $\Delta _{L,R}(\vartheta )=\Delta _{0}(T,{\bf v}_{s;L,R})\cos
2(\vartheta -\chi _{L,R})$. The presented here consideration can be also
used to consider the contact of $g$-wave superconductors or $s$-wave/$d$%
-wave contact, etc.

As we restrict ourselves to the non-self-consistent model we should
calculate the order parameter $\Delta _{0}=\Delta _{0}(T,{\bf v}_{s})$ in
the bulk $d$-wave superconductor. That is the subject of the Appendix.

In the particular case considered below in detail we have ${\bf v}_{s;L}=%
{\bf v}_{s;R}={\bf v}_{s}$ and denote $\widetilde{\omega }=\omega _{n}+i{\bf %
p}_{F}{\bf v}_{s}$, $\Omega _{L,R}=\sqrt{\widetilde{\omega }^{2}+\Delta
_{L,R}^{2}}$; in this case we obtain
\begin{equation}
g(0)=\frac{\widetilde{\omega }(\Omega _{L}+\Omega _{R})-i\eta \Delta
_{L}\Delta _{R}\sin \phi }{\Omega _{L}\Omega _{R}+\widetilde{\omega }%
^{2}+\Delta _{L}\Delta _{R}\cos \phi }.  \label{g(0)_v_s_equal}
\end{equation}
In the absence of the transport current (${\bf v}_{s}=0$) in this
expression: $\widetilde{\omega }=\omega _{n}$ \cite{AOZ}.

We should also clarify the sign of the square root in $\Omega _{L,R}$. To
make the solution (\ref{g(x)}) convergent, we should require $%
\mathop{\rm Re}%
\Omega _{L,R}>0,$which fixes the sign of the square root in $\Omega _{L,R}$
to be $%
\mathop{\rm sign}%
(\omega \cdot {\bf p}_{F}{\bf v}_{s;L,R})$. Moreover, this requirement, as
it can be shown, provides the supplementary condition on $%
\mathop{\rm Re}%
g$: $%
\mathop{\rm sign}%
(%
\mathop{\rm Re}%
g)=%
\mathop{\rm sign}%
(\omega ).$

\section{Influence of the transport current on the Josephson and spontaneous
currents at the interface}

Further we study the Josephson contact for the definite case: ${\bf v}_{sL}=%
{\bf v}_{sR}={\bf v}_{s}$ and $\chi _{L}=0$ and $\chi _{R}=\pi /4$.

For small values of $v_{s}$ (in the linear in $p_{F}v_{s}/T_{c}$
approximation) we can state the following approximate relations (which are
valid for the values of $\phi $ in the vicinity of $\pm \pi /2$):
\[
j_{J}(-{\bf v}_{s},\phi )\simeq j_{J}({\bf v}_{s},\phi ),~j_{y}(-{\bf v}%
_{s},\phi )\simeq -j_{y}({\bf v}_{s},-\phi ),
\]
and for difference $\delta j\equiv j({\bf v}_{s})-j({\bf v}_{s}=0):$%
\[
\delta j_{J}(-\phi )\simeq -\delta j_{J}(\phi ),~\delta j_{y}(-\phi )\simeq
\delta j_{y}(\phi ),
\]
while at ${\bf v}_{s}=0$%
\[
j_{J}(-\phi )=-j_{J}(\phi ),~j_{y}(-\phi )=-j_{y}(\phi ).
\]
In the linear approximation the shift current $\delta j_{y}$ is an even
function of $\phi $ in contrast to $j_{y}({\bf v}_{s}=0)$. For the
spontaneous current (at $\phi =\pm \pi /2$) the shift currents $\delta j_{y}$
are equal:
\begin{eqnarray}
j_{y}(\phi &=&\pm \pi /2)=j_{S}+\delta j_{y},  \label{j_S_linear_approx} \\
j_{S}(-\pi /2) &=&-j_{S}(\pi /2),~\delta j_{y}(-\pi /2)=\delta j_{y}(\pi /2).
\nonumber
\end{eqnarray}
In nonlinear consideration these shift currents are different for the two
cases and it is discussed below.

\begin{figure}[t]
\centering
\includegraphics[width=8 cm]{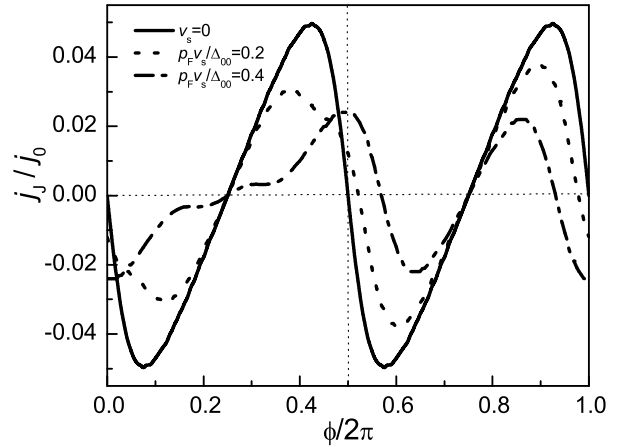}
\caption{Josephson current density through the interface $j_{J}$
versus
phase $\protect\phi $ ($\protect\chi _{L}=0,$ $\protect\chi _{R}=\protect\pi %
/4,$ $T=0.1T_{c}$); $\Delta _{00}=\Delta
_{0}(T=0,v_{s}=0)=2.14T_{c}$.} \label{j_J for d-wave}
\end{figure}

\begin{figure}[t]
\centering
\includegraphics[width=8 cm]{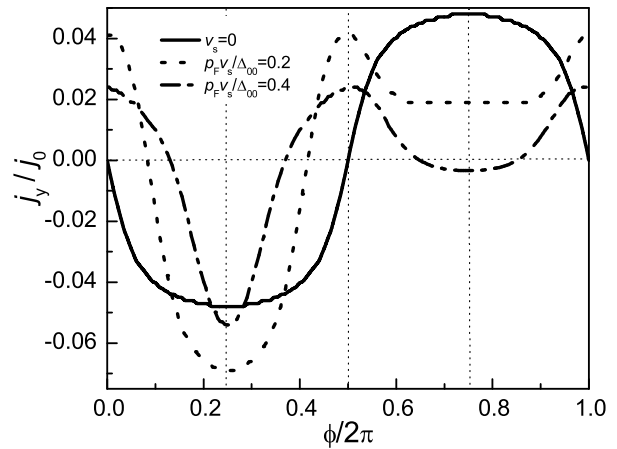}
\caption{Tangential current density at the interface $j_{y}$ versus phase $%
\protect\phi $ ($\protect\chi _{L}=0,$ $\protect\chi _{R}=\protect\pi /4,$ $%
T=0.1T_{c}$).} \label{j_S for d-wave}
\end{figure}

At Figs. \ref{j_J for d-wave}-\ref{j_S for d-wave} we plot the
normal (Josephson) and tangential components of the current
densities at the interface plane as functions of the phase
difference $\phi $ at low temperature. 

\begin{figure}[tbp] \centering
\includegraphics[width=8 cm]{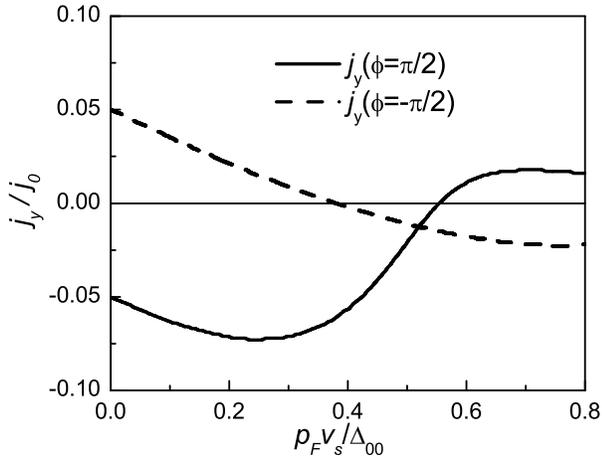}
\caption{Tangential current density at the interface $j_{y}$ for
the two values of phase difference (spontaneous current)\ versus
superfluid velocity
$v_{s}$ ($\protect\chi _{L}=0,$ $\protect\chi _{R}=\protect\pi /4,$ $%
T=0.1T_{c}$).} \label{j_S (+-Pi/2)}
\end{figure}

In the absence of the transport current: (i) ${\bf j}$ is an odd
function of $\phi $; (ii) the normal component of the current
(Josephson current) is $\pi $-periodic; (iii) in the equilibrium
state at $\phi =\pm \pi /2$: $j_{J}=0,$ $j_{y}(\pm \pi
/2)=j_{S}=\mp \left| j_{S}\right| $. In the latter case the
tangential current exists in the absence of the Josephson current;
for that reason it is referred to as the spontaneous current. The
presence of the transport current breaks (i)-(iii) symmetry
relations. There is non-zero Josephson current at $\phi =0,\pi $.
How the transport current influences the spontaneous current (i.e.
the tangential current at $\phi =\pi /2$ and $\phi =-\pi /2$) is
shown at Fig. \ref{j_S (+-Pi/2)}. 
The shift of the two values of the current for small values of $%
v_{s}$ (in the linear in $p_{F}v_{s}/T_{c}$ approximation) is equal (see Eq.
(\ref{j_S_linear_approx})); however at values $\ v_{s}\sim 0.2\Delta
_{00}/p_{F}$ the shift current (i.e. the difference $j_{y}({\bf v}%
_{s})-j_{S}({\bf v}_{s}=0)$) is of different signs for the two currents and
of opposite to $j_{S}$ directions.

We also note the following relations for ${\bf v}_{s}\neq 0$: (1) $%
j_{J}(\phi =\pi )=-j_{J}(\phi =0)\neq 0$ (the presence of the transport
current induces non-zero Josephson current in the absence of an external
phase difference); (2) $j_{J}\left( \phi =\pm \frac{\pi }{2}\right) =0,~%
\frac{dj_{J}}{d\phi }\left( \phi =\pm \frac{\pi }{2}\right) >0$ (the
transport current does not change the values of equilibrium phase
difference, at $\phi =\pm \pi /2$); (3) $j_{y}(\phi =\pi )=j_{y}(\phi
=0)\neq 0.$ The latter relation concerns the interesting phenomena, studied
in \cite{K0Sh(2003)}: for some values of phase difference (here in the
vicinity of $\phi =0,\pi $) the interference of the angle-dependent
condensate wave functions results in the appearance of the additional
tangential current with the direction opposite to the transport current in
the banks. We underline that the resulting tangential current is not a sum
of the spontaneous current and the transport current \cite{K0Sh(2003)}.
Thus, the transport current drastically influences both the tangential
(spontaneous) and Josephson currents.

We can write down explicitly simple expression for the current for
temperatures close to the critical (so close that $\Delta
_{0},p_{F}v_{s}\ll T_{c}$). From Eq. (\ref{g(0)_v_s_equal}) we
have:
\begin{eqnarray}
\mathop{\rm Im}%
g(0) &\simeq &\Delta _{L}\Delta _{R}[-\eta \frac{1}{2\omega _{n}^{2}}\sin
\phi +\frac{{\bf p}_{F}{\bf v}_{s}}{\omega _{n}^{3}}\cos \phi +
\label{Img(0)_close_to_Tc} \\
&&+\eta \frac{3}{2}\frac{({\bf p}_{F}{\bf v}_{s})^{2}}{\omega _{n}^{4}}\sin
\phi +\eta \frac{\Delta _{L}\Delta _{R}}{8\omega _{n}^{4}}\sin 2\phi ].
\nonumber
\end{eqnarray}
At $\chi _{L}=0$ and $\chi _{R}=\pi /4$ this results in the following:
\begin{eqnarray}
{\bf j} &=&{\bf j}_{J}+{\bf j}_{S}+\widetilde{{\bf j}}, \\
{\bf j}_{J} &=&-\frac{1}{3024\pi }j_{0}\frac{\Delta _{0}{}^{4}}{T_{c}^{4}}%
\sin 2\phi \cdot {\bf e}_{x}, \\
{\bf j}_{S} &=&-\frac{1}{60\pi }j_{0}\frac{\Delta _{0}{}^{2}}{T_{c}^{2}}\sin
\phi \cdot {\bf e}_{y}, \\
\widetilde{{\bf j}} &=&\frac{3}{560\pi }j_{0}\frac{\Delta _{0}{}^{2}}{%
T_{c}^{2}}\frac{(p_{F}v_{s})^{2}}{T_{c}^{2}}\sin \phi \cdot {\bf e}_{y}.
\label{j_JT}
\end{eqnarray}
Here $\Delta _{0}=\Delta _{0}(T,v_{s})$ and is defined by Eq. (\ref
{Delta_close_to_Tc}). In particular, at $v_{s}=0$ this gives:
\begin{eqnarray}
j_{J} &=&-1.7\cdot 10^{-2}j_{0}\left( 1-\frac{T}{T_{c}}\right) ^{2}\sin
2\phi ,  \label{j_J_at_T_close_to_Tc} \\
j_{S} &=&-6.6\cdot 10^{-2}j_{0}\left( 1-\frac{T}{T_{c}}\right) \sin \phi .
\label{j_S_at_T_close_to_Tc}
\end{eqnarray}

We note that $\widetilde{{\bf j}}=-\frac{9}{28}\frac{(p_{F}v_{s})^{2}}{%
T_{c}^{2}}{\bf j}_{S}$. It follows that the effect of transport current on
''spontaneous'' tangential current at $T\sim T_{c}$ is to reduce its value
by a small shift. It is remarkable that the tangential to the boundary
current contains only corrections of the second order on the parameter $%
p_{F}v_{s}/T_{c}$ \footnote{%
There is also a term with the factor $\frac{p_{F}v_{s}}{T_{c}}\frac{\Delta
_{0}^{4}}{T_{c}^{4}}$, which is neglected here. This term results in the
equal shifts of $j_{S}$ for $\phi =\pm \pi /2$.}. If $\chi _{L}=0$ and $\chi
_{R}=\delta \chi \neq \pi /4$, the integration of the second term in Eq. (%
\ref{Img(0)_close_to_Tc}) would give us the factor $\pi \cos ^{2}\delta \chi
-\frac{\pi }{2}$, which is zero for $\delta \chi =\frac{\pi }{4}$; this term
at $\delta \chi =0$\ and $\phi =0$ gives the homogeneous current density
(Eq. (\ref{j_at_T_close_to_Tc})).

The integration of the first term in Eq. (\ref{Img(0)_close_to_Tc}) gives us
the factor $\cos 2\delta \chi $ for $x$-component of the current and $\sin
2\delta \chi $ for $y$-component. In the case of $\delta \chi =\pi /4$ this
term gives only the tangential component. As the consequence $j_{S}\gg j_{J}$
(see Eqs. (\ref{j_J_at_T_close_to_Tc})-(\ref{j_S_at_T_close_to_Tc})).

It was discussed above that the linear in $p_{F}v_{s}/T_{c}$ terms result in
the uniform shift of $j_{S}$. We can see that nonlinear terms result in the
shift of different sign, and in both cases of opposite to $j_{S}$ direction
(see Eq. (\ref{j_JT})). This in part explains the nonmonotonic behavior of $%
j_{y}$ (see Fig. \ref{j_S (+-Pi/2)}). The fact, that the presence of the
transport current significantly changes the tangential (spontaneous)
currents, can be proposed to be used for its control, which is important in
view of their possible application for quantum computation \cite{Ioffe}-\cite
{Amin2003}.

\section{Conclusion}

The influence of the transport current, which flows in the banks, on the
stationary Josephson effect in the contact of two $d$-wave superconductors
is studied. We have derived the equations, which allow general consideration
of the contact of two singlet superconductors with different axes
orientation and with different transport currents in the banks.
Particularly, we have studied the planar contact of two $d$-wave
superconductors in the case of $\pi /4$-misorientation with equal transport
currents in the banks. It was demonstrated that the current-phase relation
drastically depends upon the value of the transport current. The ground
state degeneracy in the absence of transport current (at $\phi =\pm \pi /2$)
is removed at $v_{s}\neq 0$. The dependence of the shift current (which is
the difference of the tangential current and the spontaneous one) on $v_{s}$%
\ is shown to be non-linear. It is proposed to use the transport current for
the control of qubits based on the contact of two $d$-wave superconductors.

We acknowledge support from D-Wave Sys. Inc. (Vancouver).

Results of the present study were reported at International
Conferences: "Applied Electrodynamics of high-$T_{c}$
Superconductors", IRE, Kharkov, Ukraine (May 2003) and "Josephson
junctions: Basic Studies and Novel Applications", Jena, Germany
(June 2003).

\section{Appendix{\bf . }Order parameter in the homogeneous current state}

In this section we study the homogeneous current state in the bulk d-wave
superconductor (see also in \cite{FGS(1999)}). We note, that the order
parameter $\Delta _{0}$ is the function of temperature $T$, superfluid
velocity $v_{s}$, and the angle $\chi $ between the crystallographic axis $a$
and the direction of superfluid velocity ${\bf v}_{s}$. For that we should
solve Eqs.(\ref{Eq for Delta}) and (\ref{j}) with $g$ and $f$ given by Eqs. (%
\ref{g(inf)}) and (\ref{f(inf)}):
\[
\frac{1}{\lambda }=2T\sum_{\omega >0}\int\limits_{-\pi /2}^{\pi
/2}d\vartheta
\mathop{\rm Re}%
\frac{\Delta ^{2}(\vartheta )/\Delta _{0}^{2}}{\Omega },
\]
\[
\frac{{\bf j}}{j_{0}}{\bf =}-\frac{T}{\pi T_{c}}\sum_{\omega
>0}\int\limits_{-\pi /2}^{\pi /2}d\vartheta \widehat{{\bf v}}_{F}%
\mathop{\rm Im}%
\frac{\widetilde{\omega }}{\Omega }.
\]
Here $\lambda =N_{0}V_{d},$ $\widetilde{\omega }=\omega _{n}+i{\bf p}_{F}%
{\bf v}_{s}$, $\Omega =\sqrt{\widetilde{\omega }^{2}+\Delta ^{2}}$, $\Delta
(\vartheta )=\Delta _{0}(T,{\bf v}_{s})\cos 2(\vartheta -\chi )$.

For $T=0$ (substituting $\pi T\sum_{\omega }$ by the integral $\int d\omega $%
) we obtain the equations for the order parameter $\Delta _{0}$ and the
current density $j$:

\begin{equation}
\ln \left( \frac{\Delta _{00}}{\Delta _{0}}\right) =\frac{2}{\pi
}\int d\vartheta \left( \frac{\Delta (\vartheta )}{\Delta
_{0}}\right) ^{2}\ln
\left( \left| \frac{{\bf v}_{s}{\bf p}_{F}}{\Delta (\vartheta )}\right| +%
\sqrt{\left( \frac{{\bf v}_{s}{\bf p}_{F}}{\Delta (\vartheta )}\right) ^{2}-1%
}\right)  \label{Eq for Delta (T=0)}
\end{equation}
here $\Delta _{00}=\Delta _{0}(T=0,v_{s}=0)=\xi \omega
_{c}e^{-2/\lambda }$, $\xi =4e^{-1/2}$,

\begin{eqnarray}
\frac{j}{j_{0}} &=&-\frac{1}{4\pi }\frac{v_{s}p_{F}}{T_{c}}+  \label{j (T=0)}
\\
&&+\frac{1}{2\pi ^{2}}\int d\vartheta \left| \cos \vartheta \right| \sqrt{%
\left( \frac{{\bf v}_{s}{\bf p}_{F}}{T_{c}}\right) ^{2}-\left( \frac{\Delta
(\vartheta )}{T_{c}}\right) ^{2}}.  \nonumber
\end{eqnarray}

In Eqs. (\ref{Eq for Delta (T=0)}) and (\ref{j (T=0)}) the integration is
performed in the region where $\Delta (\vartheta )^{2}<({\bf v}_{s}{\bf p}%
_{F})^{2}$ for $\vartheta \in (-\frac{\pi }{2},\frac{\pi }{2})$.

\begin{figure}[tbp] \centering
\includegraphics[width=8 cm]{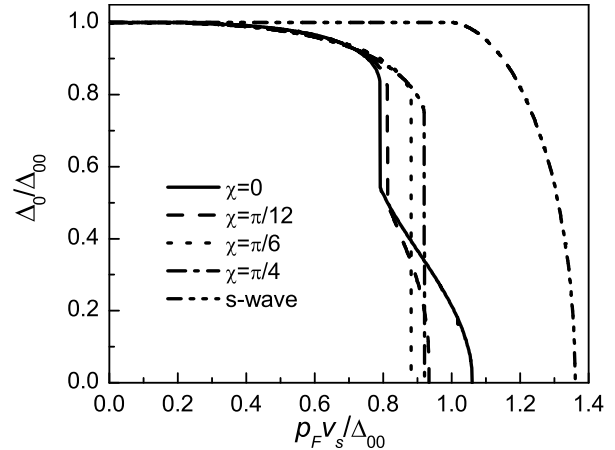}
\caption{Order parameter $\Delta _{0}(T,{\bf v}_{s})$ in the bulk
$d$-wave superconductor vs. superfluid velocity $v_{s\text{ \
}}$for different angles $\protect\chi $ between ${\bf v}_{s}$ and
${\bf a}$-axis ($T=0$).} \label{Delta(q)}
\end{figure}

\begin{figure}[tbp]
\centering
\includegraphics[width=8 cm]{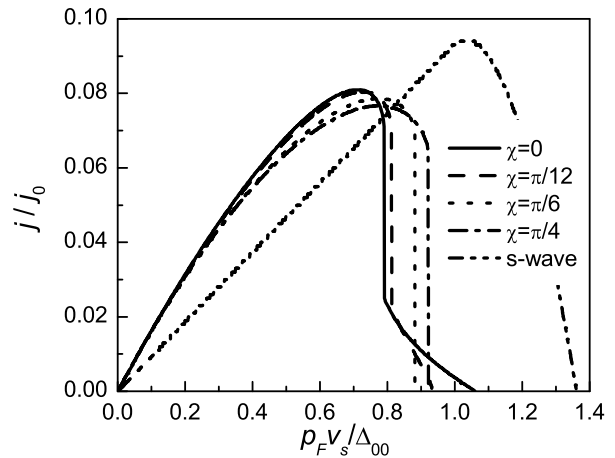}
\caption{Current density in the bulk $d$-wave superconductor vs.
superfluid velocity $v_{s\text{ \ }}$for different angles
$\protect\chi $ ($T=0$).} \label{j(q)}
\end{figure}

At Figs.\ref{Delta(q)}-\ref{j(q)} we plot the order parameter $\Delta _{0}(T,%
{\bf v}_{s})$ and the current density versus superfluid velocity
$v_{s\text{ \ }}$for different angles $\chi $ at low temperature.
For comparison we also plot the curves for the $s$-wave
superconductor. 

The numerical analysis at low temperature shows that in spite of
the strong anisotropy of the pairing
potential, the order parameter $\Delta _{0}$, the critical velocity $%
v_{s}^{cr}$ and the critical current $j_{c}$ depend weakly on the angle $%
\chi $ between ${\bf v}_{s}$ and crystallographic $a$-axis (see Figs. \ref
{Delta(q)}-\ref{j(q)} and in Ref. \cite{FGS(1999)}). Namely, the respective
difference is maximal for $\chi =0$ and $\chi =\pi /4$ and does not exceed
0.1. For small values of superfluid velocity, i.e. in the linear
approximation on the parameter $v_{s}p_{F}/T_{c}$, both $\Delta _{0}$ and $j$
do not depend on $\chi $.

For a temperature close to $T_{c}=\beta \omega _{c}e^{-2/\lambda
}=0.47\Delta _{00}$, where $\beta =\frac{2}{\pi }e^{C}$ ($C$ is
the Euler constant), both the gap function $\Delta _{0}$ and
current density $j$, do not depend upon angle $\chi $:
\begin{equation}
\Delta _{0}^{2}=\frac{32\pi ^{3}}{21\zeta (3)}T_{c}^{2}\left( 1-\frac{T}{%
T_{c}}\right) -\frac{4}{3}(p_{F}v_{s})^{2},  \label{Delta_close_to_Tc}
\end{equation}
\begin{equation}
\frac{j}{j_{0}}=-\frac{7\zeta (3)}{32\pi ^{3}}\frac{\Delta _{0}^{2}}{%
T_{c}^{2}}\frac{p_{F}v_{s}}{T_{c}}.  \label{j_at_T_close_to_Tc}
\end{equation}
The temperature dependence of critical velocity $v_{s}^{cr}$ follows from
Eq. (\ref{Delta_close_to_Tc}): $\frac{p_{F}v_{s}^{cr}}{T_{c}}=\sqrt{\frac{%
8\pi ^{2}}{7\varsigma (3)}}\sqrt{1-\frac{T}{T_{c}}}.$

\end{document}